\documentclass{elsart}

\usepackage[square,comma]{natbib}
\usepackage{graphicx}
\usepackage{pxfonts}
\usepackage{lineno}

\usepackage{amssymb}

\journal{}

\begin{document}

\thispagestyle{empty}
\begin{Large}
\textbf{DEUTSCHES ELEKTRONEN-SYNCHROTRON}

\textbf{\large{Ein Forschungszentrum der Helmholtz-Gemeinschaft}\\}
\end{Large}

DESY 12-022

February 2012

\begin{eqnarray}
\nonumber &&\cr \nonumber && \cr \nonumber &&\cr
\end{eqnarray}
\begin{eqnarray}
\nonumber
\end{eqnarray}
\begin{center}
\begin{Large}
\textbf{On quantum effects in spontaneous emission by a relativistic
electron beam in an undulator}
\end{Large}
\begin{eqnarray}
\nonumber &&\cr \nonumber && \cr
\end{eqnarray}

\begin{large}
Gianluca Geloni,
\end{large}
\textsl{\\European XFEL GmbH, Hamburg}
\begin{large}

Vitali Kocharyan and Evgeni Saldin
\end{large}
\textsl{\\Deutsches Elektronen-Synchrotron DESY, Hamburg}
\begin{eqnarray}
\nonumber
\end{eqnarray}
\begin{eqnarray}
\nonumber
\end{eqnarray}
ISSN 0418-9833
\begin{eqnarray}
\nonumber
\end{eqnarray}
\begin{large}
\textbf{NOTKESTRASSE 85 - 22607 HAMBURG}
\end{large}
\end{center}
\clearpage
\newpage

\begin{frontmatter}



\title{On quantum effects in spontaneous emission by a relativistic electron
beam in an undulator}


\author[XFEL]{Gianluca Geloni\thanksref{corr},}
\thanks[corr]{Corresponding Author. E-mail address: gianluca.geloni@xfel.eu}
\author[DESY]{Vitali Kocharyan}
\author[DESY]{and Evgeni Saldin}

\address[XFEL]{European XFEL GmbH, Hamburg, Germany}
\address[DESY]{Deutsches Elektronen-Synchrotron (DESY), Hamburg,
Germany}

\begin{abstract}
Robb and Bonifacio (2011) claimed that a previously neglected
quantum effect results in noticeable changes in the evolution of the
energy distribution associated with spontaneous emission in long
undulators. They revisited theoretical models used to describe the
emission of radiation by relativistic electrons as a continuous
diffusive process, and claimed that in the asymptotic limit for a
large number of undulator periods the evolution of the electron
energy distribution occurs as discrete energy groups according to
Poisson distribution. We show that these novel results have no
physical sense, because they are based on a one-dimensional model of
spontaneous emission and assume that electrons are sheets of charge.
However, electrons are point-like particles and, as is well-known,
the bandwidth of the angular-integrated spectrum of undulator
radiation is independent of the number of undulator periods. If we
determine the evolution of the energy distribution using a
three-dimensional theory we find the well-known results consistent
with a continuous diffusive process. The additional pedagogical
purpose of this paper is to review how quantum diffusion of electron
energy in an undulator with small undulator parameter can be simply
analyzed using the Thomson cross-section expression, unlike the
conventional treatment based on the expression for the
Lienard-Wiechert fields.
\end{abstract}

%
%
%
\end{frontmatter}



\section{\label{sec:intro} Introduction}

In a recent article \cite{BONI} it is stated that quantum effects in
spontaneous emission by a relativistic electron beam in an undulator
can be described by a drift-diffusion equation only when the
parameter

\begin{eqnarray}
\epsilon = \frac{N_w \hbar \omega}{\gamma m c^2}~,\label{eps}
\end{eqnarray}
is much smaller than unity, where $N_w$ is the number of undulator
periods, $\hbar$ is the reduced Planck constant, $\omega$ is the
photon frequency, $\gamma$ the relativistic Lorentz factor, $m$ the
electron rest mass and $c$ the speed of light. In that work it is
argued that when $\epsilon \geq 1$, a drift-diffusion equation is no
more sufficient to describe the the evolution of the distribution of
electron momenta, which "occurs as discrete momentum groups
according to a Poisson distribution".

In this paper we will show that results in \cite{BONI} are
incorrect, because they are based on a one-dimensional model of the
spontaneous radiation emission.  This model does not account for the
angular distribution of the radiation, but only for the emission on
axis, which is characterized by an overall relative bandwidth $\sim
1/N_w$. In contrast to this, the electron recoil related with the
quantized nature of photons depends on the entire angular
distribution of the radiation, which is fundamentally linked to the
Thomson scattering phenomenon in the case for a small undulator
parameter $K \ll 1$, \cite{DERB,BENS}. When the angular distribution
of radiation is properly accounted for, the overall,
angle-integrated relative bandwidth is independent on the number of
undulator periods. As a result, it turns out that a
three-dimensional drift-diffusion model is valid when the parameter

\begin{eqnarray}
\zeta = \frac{\hbar \omega}{\gamma m c^2}~,\label{zeta}
\end{eqnarray}
is much smaller than unity. This means that a Fokker-Planck approach
is always valid in all cases of practical interest.

In this work we will first review the spectral-angular
characteristics of undulator radiation. For reasons of simplicity,
from the very beginning we will consider the limit for $N_w \gg 1$
and $K\ll 1$. Our considerations can easily be applied to arbitrary
values of $N_w$ and $K$, but the choice of $N_w \gg 1$ and $K\ll 1$
easily allows one to underline the fundamental point that the
angle-integrated spectrum of radiation does not depend on the number
of undulator periods $N_w$. Using a Fokker-Planck equation we will
derive the diffusion coefficient in agreement with \cite{SAL1}.
Finally, the diffusion coefficient will also be derived by
exploiting the relation between undulator radiation and Thomson
scattering, which stresses once more the intrinsic three-dimensional
nature of the radiation pattern.

\section{Spontaneous emission process and associated quantum effects}

\subsection{Spectral-angular distribution of radiation}

As is well-known, spontaneous radiation emission from an
ultrarelativistic electron in an undulator can be modeled fully
classically as long as the energy of the emitted photons $\hbar
\omega$ is much smaller than the electron energy $\gamma m c^2$. In
this case, the knowledge of the classical characteristics of
radiation can easily be used to discuss quantum effects on the
electron motion integrated along the trajectory.

Characteristics of spontaneous radiation have been studied long time
ago in \cite{ALFE,KINC}. In this section, we briefly review them,
focusing on the particular case of a planar undulator, and following
notations introduced in previous works of us \cite{OUR1}. In order
to do so, we first call with $\vec{\bar{E}}(\omega)$ the transverse
component of the electric field generated by an electron in the
space-frequency domain\footnote{By this, $\vec{\bar{E}}(\omega)$ is
defined as the Fourier transform of the electric field in the time
domain, $\vec{{E}}(t)$, according to
$\vec{\bar{E}}(\omega)=\int_{-\infty}^{\infty} \vec{{E}}(t) \exp[i
\omega t] dt$, and has a dimension of an electric field multiplied
by a time.}. Based on the ultrarelativistic approximation $\gamma^2
\gg 1$ and on the consequent paraxial approximation, we introduce
the slowly varying electric field envelope $\vec{\widetilde{E}} =
\vec{\bar{E}} \exp{[-i\omega z/c]}$, which does not vary much along
the longitudinal coordinate $z$ on the scale of the reduced
wavelength $\lambdabar=\lambda/(2\pi)$. We can specify "how near"
$\omega$ is to the resonant frequency of the undulator,
$\omega_{r0}=2 k_w c \bar{\gamma}_z^2$,  by introducing a detuning
parameter $C$, defined as $C = \omega /(2\bar{\gamma}_z^2c)-k_w =
({\Delta\omega}/{\omega_{r0}}) k_w$, where $\omega = \omega_{r0} +
\Delta \omega$. Here $k_w = 2\pi/\lambda_w$, $\lambda_w$ is the
undulator period, $K$ is the undulator parameter, which is related
to the undulator magnetic field $B$ by

\begin{eqnarray}
B = \frac{K m c^2 k_w}{e}~, \label{Bfield}
\end{eqnarray}
and $\bar{\gamma}_z = \gamma/\sqrt{1+K^2/2}$. We further simplify
our considerations by considering from the beginning the case for
$K^2 \ll 1$ and $N_w \gg 1$. These two assumptions do not change the
nature of our considerations, and are introduces for simplicity
only. One obtains

\begin{eqnarray}
{\vec{\widetilde{E}}} &=& - \frac{\omega K e L_w}{2 c^2 z \gamma}
\exp\left[i\frac{\omega \theta^2 z}{2c}\right] \left\{\left[1-\frac{
\theta_x^2 \omega}{ k_w c}\right]\vec{e}_x +\left[\frac{ \theta_x
\theta_y \omega}{ k_w c}\right]\vec{e}_y\right\}\cr &&\times
\mathrm{sinc}\left[\frac{L_w}{4}\left(C + {\omega \theta^2 \over{2 c
}}\right) \right] ~, \label{undurad4}
\end{eqnarray}
where $\theta_x$ and $\theta_y$ are horizontal and vertical angles
identifying the angular position of an observer and
$\theta^2=\theta_x^2+\theta_y^2$. Here and everywhere in this paper
we will be using Gaussian units. The total energy emitted per unit
spectral interval per unit solid angle turns out to be

\begin{eqnarray}
\frac{d W}{d\omega d\Omega} = \frac{\omega^2  K^2 L_w^2 e^2}{16\pi^2
c^3 \gamma^2} \left\{\left[1-\frac{ \theta_x^2 \omega}{ k_w
c}\right]^2 +\left[\frac{ \theta_x \theta_y \omega}{ k_w
c}\right]^2\right\} \mathrm{sinc}^2\left[\frac{L_w}{4}\left(C +
{\omega \theta^2 \over{2 c }}\right) \right] ~,\cr && \label{dWdOdw}
\end{eqnarray}
in agreement with \cite{KINC}.

\subsection{Angle-integrated spectral distribution of radiation}

We now integrate Eq. (\ref{dWdOdw}) over all angles by using the
fact that $N_w \gg 1$. When this is the case, the bandwidth of the
radiation spectrum does not depend on the number of undulator
periods. This is represented, mathematically, by the fact that the
$\mathrm{sinc}$ function in Eq. (\ref{dWdOdw}) can be substituted
with a Dirac-$\delta$ function according to
$\mathrm{sinc}^2[x/a]/(\pi a) \longrightarrow \delta(x)$ for
$a\longrightarrow 0$.  Integrating over the solid angle we obtain

\begin{eqnarray}
\frac{d W}{d\omega} = \frac{e^2 \omega K^2 L_w}{4 c^2 \gamma^2}
\left[1+\left(\frac{\omega}{c k_w \gamma^2}-1\right)^2\right]~,
\label{dWdw}
\end{eqnarray}
for $\omega < 2 c \gamma^2 k_w$, and zero otherwise. Note that here
we already set $\bar{\gamma}_z \simeq \gamma$ in the limit for $K
\ll 1$. Eq. (\ref{dWdw}) is in agreement with expressions in
literature, e.g. \cite{ALFE} (where the energy spectrum was first
calculated) and \cite{AMIR}\footnote{A typing error is present in
Eq. (2.11) of \cite{AMIR}.}.

For us, the important point to be underlined by inspection of Eq.
(\ref{dWdw}) is the fact that the radiation spectrum depends on the
number of undulator periods only through a scaling factor. In other
words, the bandwidth is independent of $N_w$. The reason for this is
that we are now considering the spectrum integrated over angles. At
variance, the on-axis spectral bandwidth exhibits a dependence on
the number of undulator periods, and scales as $1/N_w$. The authors
of \cite{BONI} consider form the very beginning a one-dimensional
model and explicitly state that "the linewidth of wiggler radiation
is $\Delta \omega/\omega \sim 1/N_w$". This is correct if one
considers the on-axis spectrum only, for example analyzing the
undulator output through a pinhole. In our case of interest,
however, we want to discuss the effect of the electron recoil due to
the quantized nature of radiation, and the electron does not
distinguish radiation emitted on axis from radiation emitted at an
angle. The one-dimensional model in \cite{BONI} cannot be applied,
and the linewidth of the radiation is independent of $N_w$.

\subsection{Drift-diffusion model}

The previous derivations and observations should convince the reader
that the parameter $\epsilon$ defined in Eq. (\ref{eps}) is
unphysical, and that a Fokker-Planck equation can properly describe
the evolution of the electron density, as long as $\zeta = \hbar
\omega/(\gamma m c^2) \ll 1$. The coefficient of quantum diffusion
in a bending magnet was calculated for the first time in
\cite{SAND}. This expression is valid for calculations of energy
diffusion in the undulator at large values of the undulator
parameter. At arbitrary values of $K$ the quantum diffusion
coefficient was calculated in \cite{SAL1}.

Let us write the evolution equation for a particular projection of
the electron phase space as a function the energy-time variables.
Calling with $f=f(\mathcal{E}, t)$ this projection of the electron
density phase space, and with $\psi(\mathcal{E},\Delta \mathcal{E})
d\Delta \mathcal{E}$ the probability to find an electron with energy
between $\mathcal{E}$ and $\mathcal{E} + \Delta \mathcal{E}$ in the
time interval $\Delta t$, we write the evolution equation as

\begin{eqnarray}
\frac{\partial f}{\partial t} = -C_1 \frac{\partial f}{\partial
\mathcal{E}} +\frac{1}{2} C_2 \frac{\partial^2 f}{\partial
\mathcal{E}^2} \label{FP}
\end{eqnarray}
where

\begin{eqnarray}
C_1 = \frac{1}{\Delta t} \int d \Delta \mathcal{E} ~\psi(
\mathcal{E},\Delta  \mathcal{E})\Delta  \mathcal{E} \label{C1}
\end{eqnarray}
and

\begin{eqnarray}
C_2 = \frac{1}{\Delta t} \int d \Delta \mathcal{E} ~\psi(
\mathcal{E},\Delta  \mathcal{E})\Delta  \mathcal{E}^2 \label{C2}
\end{eqnarray}
Eq. (\ref{C1}) is just the rate of mean energy lost of an electron,
Eq. (\ref{C2}) gives the diffusion coefficient we are after. We
impose energy conservation by setting $\Delta \mathcal{E} = \hbar
\omega$. By noting that

\begin{eqnarray}
\frac{1}{\hbar \omega} \frac{d W}{d (\hbar \omega)} = \psi(
\mathcal{E},\Delta  \mathcal{E}) \label{prob}
\end{eqnarray}
and using Eq. (\ref{dWdw}) we obtain

\begin{eqnarray}
&&\frac{C_2}{m^2 c^4} = \frac{d \langle (\Delta \gamma)^2)}{dt} =
\frac{c}{L_w} \frac{1}{m^2 c^4} \int_0^{\infty} d\omega ~\hbar
\omega \frac{dW}{d\omega} = \frac{7}{15} r_e c \lambdabar_c K^2
k_w^3 \gamma^4 ~.\label{coeff}
\end{eqnarray}
Not surprisingly, Eq. (\ref{coeff}) is in agreement with the result
obtained in \cite{SAL1} in the limit for $K \ll 1$. Note that
despite the use of a one-dimensional model, authors of \cite{BONI}
find parametric agreement with Eq. (\ref{coeff}) in the case for
$\epsilon \ll 1$. The reason for this is that for $\epsilon \ll 1$
they use a drift-diffusion equation. They cannot recover the exact
numerical result, since they miss the contribution to the diffusion
coefficient coming from radiation emitted at angles different from
zero, due to the incorrect choice of a one-dimensional model, but
the right parameters are nevertheless present in this asymptote.
However, the use of the one-dimensional model leads to the
introduction of the unphysical parameter $\epsilon$, and to the
consequent introduction of an artificial quantum effect that does
not exist in reality for $\epsilon \geq 1$.

\subsection{Relation with Thomson scattering}

It is straightforward to underline the well-known equivalence
between the previously obtained results and Thomson scattering of
radiation. In fact, in the limit for $K^2 \ll 1$ and $N_w \gg 1$ and
in the reference system of the electron, the undulator magnetic
field is seen as a plane wave interacting with the electron with
frequency

\begin{eqnarray}
\omega_R = \gamma c k_w ~.\label{Rest}
\end{eqnarray}
This simple observation includes the essence of the
Weizs\"{a}cker-Williams method of virtual quanta \cite{JACK}, and
allows to calculate the quantum diffusion coefficient, Eq.
(\ref{coeff}), following an alternative derivation in the rest
frame.

Due to the presence of the electron, the plane wave scatters
radiation as a function of the rest frame angle. Under the
approximation $\hbar \omega_R \ll m_e c^2$ the process differential
cross-section for horizontally polarized incident radiation is just
the Thomson cross-section for polarized radiation:

\begin{eqnarray}
\frac{d\sigma}{d\Omega_R} = r_e^2 \left[\cos^2(\theta_R) \cos^2
(\phi_R) + \sin^2(\phi_R)\right]~, \label{thompol}
\end{eqnarray}
where $\theta_R$ and $\phi_R$ are spherical coordinate angles in the
rest frame\footnote{We will label the rest frame wit $R$, and the
lab frame with $L$.} of the electron and $r_e = e^2/(m c^2)$ is the
classical electron radius.

In the language of photons we can say that Eq. (\ref{thompol}) is
related to the probability of scattering a photon in the solid angle
$d\Omega_R = \sin \theta_R d\theta_R d\phi_R$. In fact, remembering
that the radiation pulse in the rest frame has a duration given by
$L_w/(\gamma c)$, the number of photons scattered in $d \Omega_R$
can be written as

\begin{eqnarray}
\frac{d N_{phR}}{d \Omega_R} = \frac{d \sigma}{d \Omega_R}
\frac{L_w}{\gamma c}\frac{1}{\hbar \omega_R}\bar{S}_R~,
\label{restI}
\end{eqnarray}
where $\bar{S}_R$ is the time-averaged Poynting vector of the
radiation incident on the electron in the rest frame. Note that only
elastic scattering takes place under the over-mentioned assumption
$\hbar \omega_R \ll m_e c^2$. Therefore, there is no change of
photon frequency in the scattering process. Since the wave packet
incident on the electron includes $N_w \gg 1$ period we can assume,
with accuracy $1/N_w \ll 1$, that the incoming wave packet is
composed of a single frequency. This explain why Eq. (\ref{restI})
is not analyzed in frequency.

The magnitude of the time-averaged Poynting vector of the radiation
incident on the electron in the rest frame can be found remembering
that\footnote{Note that $E_L = \gamma \vec{\beta} \times \vec{B}_R$,
which presents a correction of order $1/\gamma^2$ with respect to
Eq. (\ref{trans2}). In our case, this correction can be omitted and
$E_R \simeq \gamma B_L$.}

\begin{eqnarray}
&& E_R \simeq \gamma B_L \cr && B_R = \gamma B_L \label{trans2}
\end{eqnarray}
where $B_L$ is the undulator field in the laboratory frame. With the
help of Eq. (\ref{Bfield}), the time-averaged Poynting vector of the
radiation incident on the electron in the rest frame can be written
as

\begin{eqnarray}
\bar{S}_R = \frac{c}{8 \pi} \left(\frac{\gamma K m c^2
k_w}{e}\right)^2 ~.\label{SR}
\end{eqnarray}
The relation between frequencies in the laboratory frame and in the
rest frame obey the following Lorentz transformation

\begin{eqnarray}
\omega_L(\theta_R) = \gamma \omega_R (1+\cos\theta_R)~. \label{DPL}
\end{eqnarray}
By energy conservation we can identify the change in the electron
energy in the laboratory frame with the photon energy in the
laboratory frame. Eq. (\ref{DPL}) allows us to calculate this
quantity by averaging over the number of photons scattered at angles
$\theta_R$ in the rest frame, Eq. (\ref{restI}). We can write the
rate of change in the spread of $\Delta \gamma$ as

\begin{eqnarray}
\frac{d \langle (\Delta \gamma)^2)}{dt} = \frac{c}{L_w}  \int
d\Omega_R ~ \left(\frac{\hbar \omega_L(\theta_R)}{m c^2}\right)^2
\frac{d N_{phR}}{d \Omega_R} ~.\label{coeff2}
\end{eqnarray}
With the help of Eqs. (\ref{Rest})-(\ref{restI}), Eq. (\ref{SR}) and
Eq. (\ref{DPL}) we find

\begin{eqnarray}
\frac{d \langle (\Delta \gamma)^2)}{dt} = \frac{7}{15} r_e c
\lambdabar_c K^2 k_w^3 \gamma^4~,\label{coeffend}
\end{eqnarray}
in perfect agreement with Eq. (\ref{coeff}).

\section{Conclusions}

In this paper we showed that quantum effects in spontaneous
radiation emission can be satisfactorily modeled via a
drift-diffusion model. It is of fundamental importance to treat
spontaneous radiation within a three-dimensional model. This is
explained by the fact that an electron feeling photon recoil does
not filter photons along a privileged direction, but reacts to
photons emitted at all angles. In this case, contrarily to what has
been argued in \cite{BONI}, the linewidth of spontaneous radiation
must be integrated over all angles, and is independent of the number
of undulator periods $N_w$. It follows from our analysis that if one
enforces a three-dimensional model for the spontaneous emission, a
drift-diffusion model remains valid up to photon energies smaller
than the electron energy, which practically means always. This
conclusion is also in contrast with \cite{BONI}, where the
assumption of a linewidth scaling with $1/N_w$ leads to the
identification of an unphysical parameter scaling as $N_w$, and to
the rise of artificial quantum effects when this parameters becomes
comparable with unity.

\section{Acknowledgements}

We are grateful to Massimo Altarelli, Reinhard Brinkmann,
Serguei Molodtsov and Edgar Weckert for their support and their interest during the compilation of this work.

\end{document}